\documentclass[pra,twocolumn,superscriptaddress]{revtex4}

\usepackage{times,amsmath,amsfonts,amssymb}
\usepackage{epstopdf}
\usepackage{graphicx}
\usepackage{float}

\usepackage[usenames]{color}
\usepackage{epsf}

\newcommand{\li}[1]{#1}
\newcommand{\marton}[1]{#1}
\definecolor{darkgreen}{rgb}{0.1, 0.7, 0.3}
\definecolor{purple}{rgb}{0.7, 0.1, 0.3}

\newcommand{\daniel}[1]{#1}

\newcommand{\beq}{\begin{equation}}
\newcommand{\eeq}{\end{equation}}
\newcommand{\ket}[1]{\left\vert#1\right\rangle}
\newcommand{\bra}[1]{\left\langle#1\right\vert}

\renewcommand{\phi}{\varphi}
\renewcommand{\epsilon}{\varepsilon}

\DeclareFontFamily{U}{euc}{}
\DeclareFontShape{U}{euc}{m}{n}{<-6>eurm5<6-8>eurm7<8->eurm10}{}
\DeclareSymbolFont{AMSc}{U}{euc}{m}{n}
\DeclareMathSymbol{\umu}{\mathord}{AMSc}{"16}

\begin{document}
\title{Quantum correlations at infinite temperature: the dynamical Nagaoka effect}

\author{M\'arton Kan\'asz-Nagy} 
\affiliation{Department of Physics, Harvard University, Cambridge MA 02138, United States}
\author{Izabella Lovas} 
\affiliation{MTA-BME Exotic Quantum Phases "Momentum" Research Group and Department of Theoretical Physics, Budapest University of Technology and Economics, 1111 Budapest, Hungary}
\author{Fabian Grusdt} 
\affiliation{Department of Physics, Harvard University, Cambridge MA 02138, United States}
\author{Daniel Greif} 
\affiliation{Department of Physics, Harvard University, Cambridge MA 02138, United States}
\author{Markus Greiner} 
\affiliation{Department of Physics, Harvard University, Cambridge MA 02138, United States}
\author{Eugene A. Demler}
\affiliation{Department of Physics, Harvard University, Cambridge MA 02138, United States}

\begin{abstract}
Do quantum correlations play a role in high temperature dynamics of many-body systems? A common expectation is that thermal fluctuations lead to fast decoherence \marton{ and make dynamics classical}. In this paper, we provide a striking example of a single particle \li{created} in a featureless, infinite temperature spin bath which \marton{not only exhibits non-classical dynamics but also induces strong long-lived} correlations between \marton{the surrounding spins}. We study the non-equilibrium dynamics of a hole created in \marton{a fermionic or bosonic Mott insulator} in the atomic limit, which corresponds to a degenerate spin system. In the absence of interactions, the spin correlations arise purely from quantum interference, \marton{and the correlations are both antiferromagnetic and ferromagnetic, in} striking contrast to the equilibrium Nagaoka effect. These results are relevant for several condensed matter spin systems, and should be observable using state of the art bosonic or fermionic quantum gas microscopes.
\end{abstract}

\date{\normalsize{\today}}

\maketitle

\begin{figure*}[th]
\begin{center}
\includegraphics[width = 13.5cm]{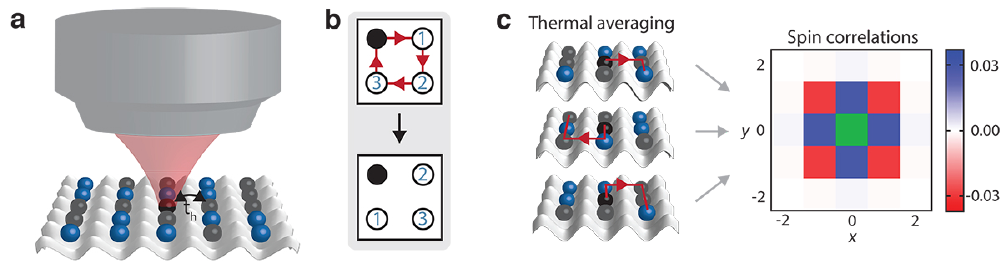}
\caption{{\bf Physical realization.} {\bf a,} Proposed experimental procedure. The system of non-interacting spins is realized by creating a Mott insulator of spinful atoms (black and blue dots) in a deep optical lattice. In the limit of infinitely strong on-site repulsion, the spin interaction vanishes. The hole (black) is created at the beginning of the experiment by removing one of the atoms; its position and the spin correlations in the environment can be measured after a propagation time $t$ using a quantum gas microscope. 
{\bf b,} As the hole moves along the trajectory indicated by arrows (top) the spins on this path are reordered (bottom).
{\bf c,} Spin correlations are calculated by thermal averaging over all possible spin configurations. During its dynamics, the hole explores all alternative paths simultaneously. Due to its interplay with the spins, the hole permutes each of these environments differently, leading to non-vanishing spin correlations even after thermal averaging.
Right panel: Spin correlations $C_{o l}$ \daniel{in the laboratory frame} between the origin $o$ (green site) and site $l$, whose coordinates are denoted by $x$ and $y$. The calculations were performed at time $t=1.1$  in a spin $S=1/2$ environment.
}
\label{fig:holeprop}
\end{center}
\end{figure*}

\noindent
Understanding the role of quantum coherence in dynamics of many-body systems at high temperatures remains a challenging open problem. Usually coherence is fragile and quickly destroyed by interaction with the environment and by local fluctuations inherent to thermal ensembles. Hence it is commonly assumed that observing quantum coherent dynamics requires preparing isolated quantum systems close to their ground states. Famous experimental demonstrations of quantum coherence, including interference of Cooper pairs in nanostructures~\cite{Friedman2000QuantumInterference, Hofstetter2009CooperPairSplitting} or interference of atomic Bose-Einstein condensates and superfluids~\cite{Andrews1997InterferenceOfBECs, Simmonds2001QuantumInterferenceOfHe3, Chin2006SuperfluidInterference}, have all been achieved under these conditions. On the other hand it has been argued that quantum interference can lead to strong deviations from simple classical dynamics. \marton{For example, the breakdown of spin diffusion was predicted for the Heisenberg model even at infinite temperature~\cite{Hubbard1970SpinDiffusion, ChertkovKolokov1994SpinDiffusion, Lovesey1994SpinDiffusion}}. Several important examples can be found in biophysics: in photosynthesis the interplay of quantum interference and decoherence leads to a much faster energy transport than would be possible classically~\cite{Engel2007Photosynthesis, Collini2010Photosynthesis, Panitchayangkoon2010Photosynthesis};
quantum coherence has also been suggested to play a crucial role in bird navigation~\cite{Ritz2000BirdNavigation} 
and the chemistry of smelling~\cite{Turin1996Smelling}.
Understanding how quantum interference can operate at high temperatures is therefore a crucial question, with tremendous potential for quantum information science~\cite{NielsenChuang2002QuantumInfoBook, Prokofev2006DecoherenceAndQuantumWalks, Morello2006DecoherenceInSpinQubitNetworks}, condensed matter~\cite{Novoselov2007RoomTemperatureQuantumHall} and biology~\cite{SethLloyd2011QuantumCoherenceInBiologicalSystems}. 

Whereas it is well understood that the entanglement of a subsystem with its environment leads to dephasing that drives the subsystem towards classical behavior, the fate of quantum coherence created in the environment is much less discussed. It is conventionally assumed that the environment's coherence quickly vanishes due to dephasing among its large number of degrees of freedom~\cite{Zurek1991Decoherence}. Here, we show however that this is not necessarily the case. We present a surprising example, where adding a single quantum particle to an infinite temperature spin environment can lead to appreciable dynamical correlations among the spins (Fig.~\ref{fig:holeprop}). We consider a system of non-interacting spins on a two-dimensional lattice, which is routinely realizable with bosonic or fermionic ultracold quantum gas microscopes. \daniel{In a deep optical lattice, on-site repulsion brings the atoms into a Mott state, where each site is occupied by exactly one atom. The spins are represented by internal degrees of freedom of the atoms, such as their hyperfine states~\cite{Jordens2008FermionicMI} or nuclear spins~\cite{BlochSUN}, with $\mathcal{N}=2S+1$ degrees of freedom, modeling a spin $S$ system. In the limit of strong on-site repulsion, the spins completely decouple, as virtual tunneling to the neighboring sites is suppressed. This realizes a non-interacting spin system discussed here.} Removing a spin on one site creates a hole that can move on the lattice at no energy cost and permute the spins during its motion~\cite{Nagaoka1965Ferromagnetism, Nagaoka1966Ferromagnetism, Auerbach2012Book, Carlstrom2016HolePropagation} (Fig.~\ref{fig:holeprop}~a-b). In contrast to a classical particle performing Brownian motion, that would only scramble the random spins along its path and keep the environment completely disordered, the quantum mechanical hole is capable of exploring alternative paths in parallel. As each path can lead to different permutations of the spins, the superposition of these outcomes creates entanglement in the spin bath. This leads to dynamical spin correlations in the environment, whereas individual sites remain paramagnetic. In contrast to the usual polaron effect, where a particle locally modifies its environment due to their interaction~\cite{Landau1933Polaron, Frohlich1954Polaron, 
Mishchenko2000DiagrammaticQMC, Zhang1991tJHolePropagation,  Mishchenko2001tJModel, White2001tJPolaronDMRG, Shchadilova2016QuantumDynamicsOfUltracoldBosePolarons, Jorgensen2016ObservationOfAttractiveAndRepulsivePolarons, Hu2016BosePolaronsInTheStronglyInteractingRegime}, 
these correlations arise purely from quantum interference. 

\begin{figure*}[th]
\begin{center}
\includegraphics[width = 14cm]{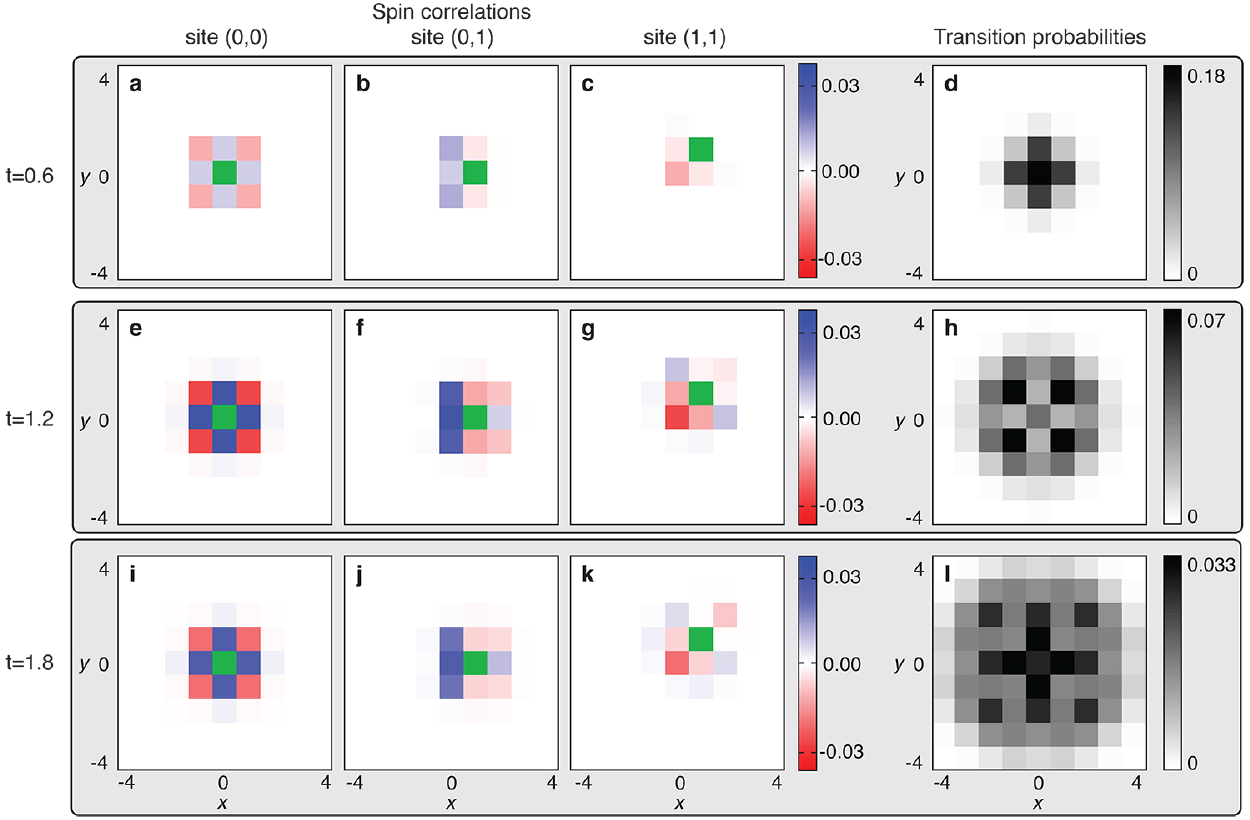}
\caption{{\bf Spatial spin correlations.} Induced spin correlations of the hole in a degenerate spin $S=1/2$ environment \daniel{in the laboratory frame} (left). The reference site $j$ is denoted by green, whereas $x$ and $y$ specify the coordinates of the second site $l$. The reference site is chosen to be $(0,0)$ for {\bf a}, {\bf e} and {\bf i}; $(0,1)$ for {\bf b}, {\bf f} and {\bf j}; and $(1,1)$ for  {\bf c}, {\bf g} and {\bf k}. The probability density of the hole, also discussed in Ref.~\cite{Carlstrom2016HolePropagation}, is exhibited on the right.  Results are shown at times $t=0.6$ ({\bf a}-{\bf d}), $t=1.2$ ({\bf e}-{\bf h}) and $t=1.8$ ({\bf i}-{\bf l}) in a spin $S=1/2$ system.
}
\label{fig:2Dspincorr}
\end{center}
\end{figure*}

Our system is also closely related to the ideas of dissipationless decoherence~\cite{Mozyrsky1998AdiabaticDecoherence, Gangopadhyay2001DissipationlessDecoherence}, 
studied in the context of quantum information~\cite{Morello2006DecoherenceInSpinQubitNetworks, Prokofev2006DecoherenceAndQuantumWalks}, condensed matter~\cite{Prokof2000TheoryOfSpinBath}
and cosmology~\cite{Unruh2012DecoherenceWithoutDissipation}. Even though there is no energy transfer between the hole and the environment, the hole's propagation is slowed down as quantum coherence is suppressed due to its entanglement with the surrounding spins \daniel{as was studied in Ref.}~\onlinecite{Carlstrom2016HolePropagation}. The new insight of our work is that this process also induces spin correlations in the environment \daniel{(Figs.~\ref{fig:holeprop} and \ref{fig:2Dspincorr})}, and therefore these correlations and decoherence are intimately related. The non-interacting system discussed here is special in the sense that the degrees of freedom in the environment are all degenerate, which suppresses the effects of dephasing. In fact, during the time scale of our simulations, these correlations remain finite.

We identify the interference terms that make the hole's dynamics dependent on the environment's spin $S$~\cite{Carlstrom2016HolePropagation}. These terms are identical to those that generate spin correlations in the environment, and they vanish exponentially in environments of large spin. Finally, we find that the simple analytical model of a hole on the Bethe lattice~\cite{Ostilli2012CayleyTreesAndBetheLattices} closely approximates the hole's dynamics in a $S\to \infty$ environment within the time scales of the simulation. \marton{It has been suggested} that the dynamics of the hole should cross over from the initial ballistic to diffusive behavior at long times~\cite{Carlstrom2016HolePropagation}. However, this question remained inconclusive due to the limited time available \marton{for numerical simulations}. The correspondence with the Bethe lattice provides further evidence that the hole's dynamics \li{indeed crosses over to diffusive behavior}. \\

\noindent {\bf Results}\\
Dynamics of charge carriers in fluctuating and disordered spin background lies at the heart of many physical systems, including high-temperature superconductors~\cite{WenLee2006HighTcReview, Sachdev2007Book, Auerbach2012Book}, 
the paramagnetic phase of supersolid ${^3}{\rm He}$~\cite{SokoloffWidom1975MagneticOrderingInHe3, Montambaux1982He3Vacancies, Kumar1985LowTSusceptibilityInSolidHe3, Kumar1987MagneticPolaronsInHe3}, organic materials~\marton{\cite{Lebed2008OrganicSuperconductorsBook}}, \marton{manganites exhibiting colossal magneto resistance effect}~\cite{Tokura1999ColossalMagnetoresistanceManganites, Roder1996ColossalMagnetoresistanceManganites}, and \daniel{multicomponent} ultracold atoms in optical lattices~\cite{Greiner2002BosonicMI, Jordens2008FermionicMI, Greif2016FermionicMI}. \daniel{The Hubbard model provides a paradigmatic model of these systems, characterized by the nearest neighbor tunneling energy $t_{\rm h}$ and an on-site repulsion between the atoms.} As the spinful atoms or electrons in each of these systems repel each other strongly, they occupy individual lattice sites, \daniel{realizing} a Mott insulator of spins~\cite{Auerbach2012Book, Greiner2002BosonicMI, Jordens2008FermionicMI, Greif2016FermionicMI}. Assuming spin-independent on-site \daniel{repulsion $U$, a spin interaction $J$} of the order of $t_{\rm h}^2/U$ is provided by virtual tunneling to neighboring sites, \daniel{leading to the so-called $t-J$ model~\cite{Auerbach2012Book}. The spin coupling $J$ then vanishes in the limit of large on-site interactions $U \to \infty$, realizing the non-interacting spin system studied here.}

Despite its simplicity, the degenerate spin environment has surprisingly rich physics. As has been shown by Nagaoka~\cite{Nagaoka1965Ferromagnetism, Nagaoka1966Ferromagnetism}, the ground state of the system becomes ferromagnetically ordered in the presence of a single hole, as this state provides free propagation to the hole so that it can minimize its kinetic energy. Here, we discuss the opposite limit of an infinite temperature spin environment, where the hole creates dynamical correlations among the spins. These correlations are of similar origin as the equilibrium Nagaoka effect, as they arise from the dependence of the hole's dynamics on the surrounding spin configurations: locally ferromagnetic spin domains lead to enhanced quantum coherence and to faster propagation. As the hole acts on the spins in each spin background differently, the resulting correlations are not averaged out to zero due to thermal fluctuations. However, in contrast to the Nagaoka ground state, the correlations studied in this paper are both ferromagnetic and antiferromagnetic. \\

\noindent {\bf Physical realization.}
Fig.~\ref{fig:holeprop}~a shows a possible experimental realization of our proposal. The non-interacting spin system is realized by creating a Mott insulator of fermionic or bosonis atoms in a deep optical lattice, with a single atom per site. Tuning the lattice depth allows one to reach the limit of strong on-site repulsion $U \gg t_{\rm h}$ such that the spin interactions become negligible. The hole can be created by removing a single atom at at the origin $o$, with coordinates $(0,0)$, using a quantum gas microscope that can optically address sites independently~\cite{Weitenberg2011SingleSpinAddressing}. The microscope can also measure the hole's position as well as the spin state at each site after a propagation time $t$. In order to account for thermal fluctuations at infinite temperature, this procedure has to be repeated many times, in each case with a different, random initial spin configuration, resulting in an averaging over all possible spin states, as we show in Fig.~\ref{fig:holeprop}~c.

The dynamics of the hole is governed by the Hamiltonian $\hat{H} = - t_{\rm h} \sum_{\langle j l \rangle} \hat{c}_j^\dagger  \, \hat{\mathcal{P}}_{jl} \,  \hat{c}_l$, where the operator $\hat{c}_l$ annihilates the hole at site $l$. As the hole moves from site $l$ to $j$, the operator $\hat{\mathcal{P}}_{jl}$ moves the spin at site $j$ to site $l$. Since there is no energy cost of moving the spins around, the tunneling $t_{\rm h}$ is the single energy scale of the model, and it is chosen to be $t_{\rm h} \equiv 1$, which also determines the time scale of the dynamics. After a propagation time $t$, the probability of finding the hole at site $j$ is given by $p_j(t) = \langle \hat{c}_j^\dagger \hat{c}_j \rangle (t)$. Here, the non-equilibrium average denotes $\langle \dots \rangle(t) = \frac{1}{\mathcal{N}^{M-1}} \, {\rm Tr}(\hat{c}_o e^{i \hat{H} t} \dots e^{- i \hat{H} t} \hat{c}_o^\dagger)$, where the trace sums over all possible spin configurations ${\rm Tr}(\dots) = \sum_{\Gamma} \bra{\Gamma} \dots \ket{\Gamma}$ and the denominator accounts for the of number spin configurations in the environment of $M$ sites.

Whereas the spin environment modifies the propagation of the hole~\cite{Carlstrom2016HolePropagation}, the effect of the hole on the environment can also be observed in the form of dynamical spin correlations, that are the primary focus of this work. The correlations between sites $j$ and $l$ are defined as $C_{jl}(t) = \frac{1}{S^2} \, \langle \hat{S}_j^z \hat{S}_l^z \rangle (t)$, where $\hat{S}_j^z$ denotes the $z$ component of the spin at that site, and it evaluates to $0$ when the hole is at site $j$. 
In the initial state, off-diagonal spin correlations $C_{j \neq l}$ are averaged out to zero by thermal fluctuations. Fig.~\ref{fig:2Dspincorr} shows how the introduction of the hole leads to dynamical correlations at longer times, reaching values as large as $4\%$ near the origin in a system of $S=1/2$ spins. These correlations appear as the hole extends over the lattice, so that it can build up coherence between the spins surrounding it. In the non-interacting environment, the correlations remain finite at the times available to our simulations. Since the hole cannot create spin flips, the $z$ component of the total spin of the lattice is conserved. This leads to the conservation of the sum of off-diagonal correalations (Methods) 
\beq
\sum_{j \ne l} C_{jl}(t) = 0.
\eeq
Therefore, the appearance of ferromagnetic correlations always need to be accompanied with antiferromagnetic ones and vice versa. 

The onset of spin correlations can be understood as follows. In each possible spin background, the hole permutes the spins slightly differently during its dynamics. For instance, locally ferromagnetic environments lead to slightly faster propagation due to interference terms: as the hole has no effect on ferromagnetically aligned spins, any pair of paths interfere. Spin correlations therefore evolve differently in time in each spin environment, and they are not averaged out by thermal fluctuations. 
Although an experimental realization of the infinite temperature spin background would involve averaging over all initial spin configurations, we estimate that the spin correlations can be observable in existing experimental setups~\cite{Endres2011StringOrder, Greif2016FermionicMI, Cheuk2016FermionicMIWithSingleSiteResolution, Hilker2017HiddenAFM} 
already after a few hundred measurements with a good signal to noise ratio (Supplementary Material). 
\\

{\bf Quantum interference between paths.} 
Similarly to the famous double-slit experiment~\cite{Feynman1966LecturesOnPhysics}, the probability of finding the hole at any site is determined from interference between different paths. This leads to interference fringes in the probability density of the hole~\cite{Carlstrom2016HolePropagation}. The hole's dynamics can be represented in terms of these paths by expanding its time evolution $e^{-i \hat{H} t} = \sum_{n=0}^\infty \frac{(-i \, t)^n}{n!} \, \hat{H}^n$~\cite{Nagaoka1965Ferromagnetism, Nagaoka1966Ferromagnetism}. 
Each power of $\hat{H}$ generates a step of the hole to one of its $z=4$ neighboring sites. Therefore $\hat{H}^n$ corresponds to a collection of $z^n$ possible hole paths of total length $n$. During its time evolution, the hole is in the superposition state of all paths. Since the expectation value $\langle \dots \rangle (t)$ of experimental observables, contains both the time evolution operator and its conjugate, we need to expand both of these operators in terms of paths of the hole. These are referred to as forward and backward time evolution paths, respectively. We determine the transition probability $p_j(t)$ by summing over interference terms between all pairs of forward ($\alpha$) and backward ($\beta$) evolution paths ending at site $j$. In order that two paths can interfere in a given spin environment, the hole needs to end up at the same site along both paths, and they need to produce the same final spin state. As the hole moves along these paths, it generates the permutations $\hat{\pi}_\alpha$ and $\hat{\pi}_\beta$ on the spins.
Thus, the transition probability to site $j$ is given by $p_j(t) = \sum_{\alpha, \beta} \frac{(-i \, t)^{n_\beta} \, (i \, t)^{n_\alpha}}{n_\beta ! \, n_\alpha !} \; \langle \hat{\pi}_\beta^\dagger \hat{\pi}_\alpha \rangle_0$, where the average denotes $\langle \dots \rangle_0 \equiv \langle \dots \rangle (t=0)$, and $n_\alpha$ and $n_\beta$ refer to the lengths of the paths $\alpha$ and $\beta$. The interference term between paths $\alpha$ and $\beta$ is thus depetermined by the combined permutation $\hat{\pi}_\beta^\dagger \hat{\pi}_\alpha = \hat{\pi}_\beta^{-1} \; \hat{\pi}_\alpha $ that is generated by the hole moving forward on path $\alpha$ to site $j$, and then returning to the origin on $\beta$. Due to the degeneracy of the spin environment, time-dependent observables cannot be evaluated using ordinary perturbation theory up to finite order in the hopping~\cite{Esterling1970DegenerateSpinEnvironment} (Supplementary Material). We therefore model the hole's dynamics by sampling its paths using a real-time quantum Monte Carlo algorithm~\cite{Handscomb1962MonteCarloBook, Carlstrom2016HolePropagation}. In order to account for the $\frac{t^n}{n!}$ expansion parameter and the large phase space consisting of $z^n$ paths we choose random walk paths of length $n$ from the Poisson distribution $\mathbb{P}_n \propto \frac{(z t)^n}{n!}$~\cite{Carlstrom2016HolePropagation} (Supplementary Material). The permutations generated by these paths are stored together with the acquired phase factors $i^n$ and we take all pairs of these paths to evaluate their contributions to the transition probabilities and the spin correlations (Methods).We evaluate interference terms between paths by calculating the thermal average $\langle \dots \rangle_0$ over all spin states exactly. This allows us to determine the spin correlations to high numerical accuracy, in contrast to earlier approaches~\cite{Carlstrom2016HolePropagation}.

\begin{figure*}[th]
\begin{center}
\includegraphics[width = 15.5cm]{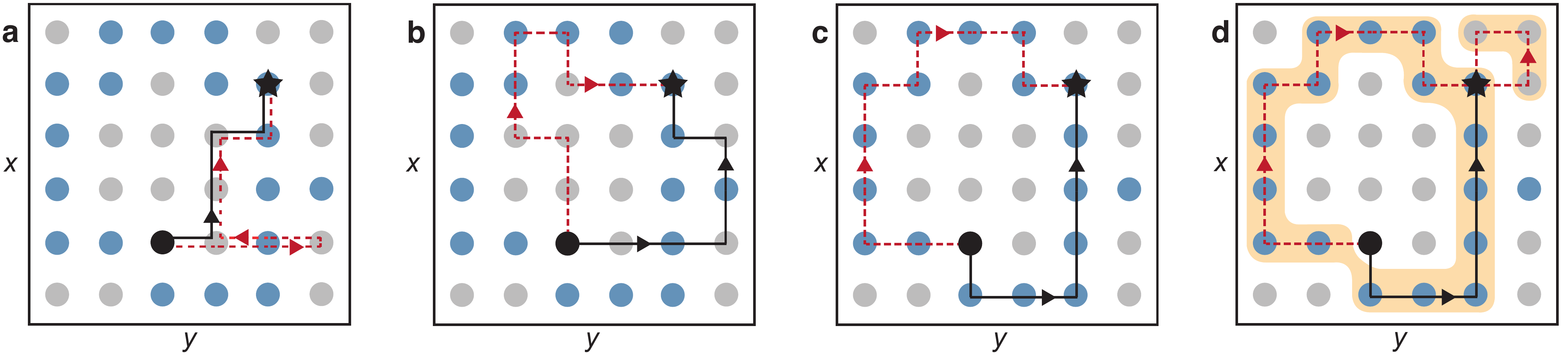}
\caption{ {\bf Interference terms.} Interference of different pairs of paths in the same initial spin background (black and blue dots). {\bf a,} By definition, the two equivalent paths (black full and red dashed lines) permute any spin state identically, making the final spin state the same. Therefore, these paths interfere in any spin background. The inequivalent paths shown in {\bf b} however bring the spin configuration into orthogonal final states, therefore their interference vanishes. In contrast, the paths in {\bf c} and {\bf d} lead to the same final state as these paths permute the spins over locally ferromagnetic regions. The dashed region in {\bf d} shows the permutation cycles generated by the hole moving along the paths. Spin correlations arise from interference between inequivalent paths {\bf b}-{\bf d}. However, these contributions vanish in environments of large spin $S\to \infty$. In these systems, the hole's dynamics is determined only by interference between equivalent paths {\bf a}.
}
\label{fig:interference}
\end{center}
\end{figure*}

The interference contributions between two paths strongly depend on how the spins are permuted as the hole moves along them. Paths that generate the same permutation of the spin environment $\hat{\pi}_\beta = \hat{\pi}_\alpha$, are referred to as being equivalent. These paths restore the original spin configuration at the end of the combined path $\hat{\pi}_\beta^\dagger \hat{\pi}_\alpha = 1$ irrespective of the spin background, leading to maximal interference  $\langle \hat{\pi}_\beta^\dagger \hat{\pi}_\alpha \rangle_0 = 1$. For example, two paths that only differ in self-retracing components are equivalent~\cite{BrinkmanRice1970}, as we show in Fig.~\ref{fig:interference}~a. However, more complicated scenarios are also possible. The path traversing a two-by-two plaquette three times is equivalent to the trivial path, where the hole stays at the origin~\cite{Trugman1988}.

Importantly, equivalent paths do not contribute to spin correlations. As they perform the same transformation on the lattice spins, thermal averaging makes the spin correlators vanish. Instead, spin correlations between lattice spins arise from pairs of inequivalent paths, that have different effect on the spins, $\hat{\pi}_\beta \ne \hat{\pi}_\alpha$. In these pairs, the combined forward and backward paths always contain loops, such as those shown in Fig.~\ref{fig:interference}~b-d. Depending on the initial spin state, the paths in these pairs often create orthogonal final spin configurations (Fig.~\ref{fig:interference}~b). Inequivalent paths can interfere only in specific initial spin states where $\hat{\pi}_\beta^\dagger \hat{\pi}_\alpha$ acts over locally ferromagnetic domains that are restored by the combined permutation (Fig.~\ref{fig:interference}~c-d). These terms thus make the hole's propagation depend on the spin state of the lattice.  Similar to the equilibrium Nagaoka effect, the correlations thus arise from the enhancement of interference terms in ferromagnetic spin domains. \\

\begin{figure}[b]
\begin{center}
\includegraphics[width = 8cm]{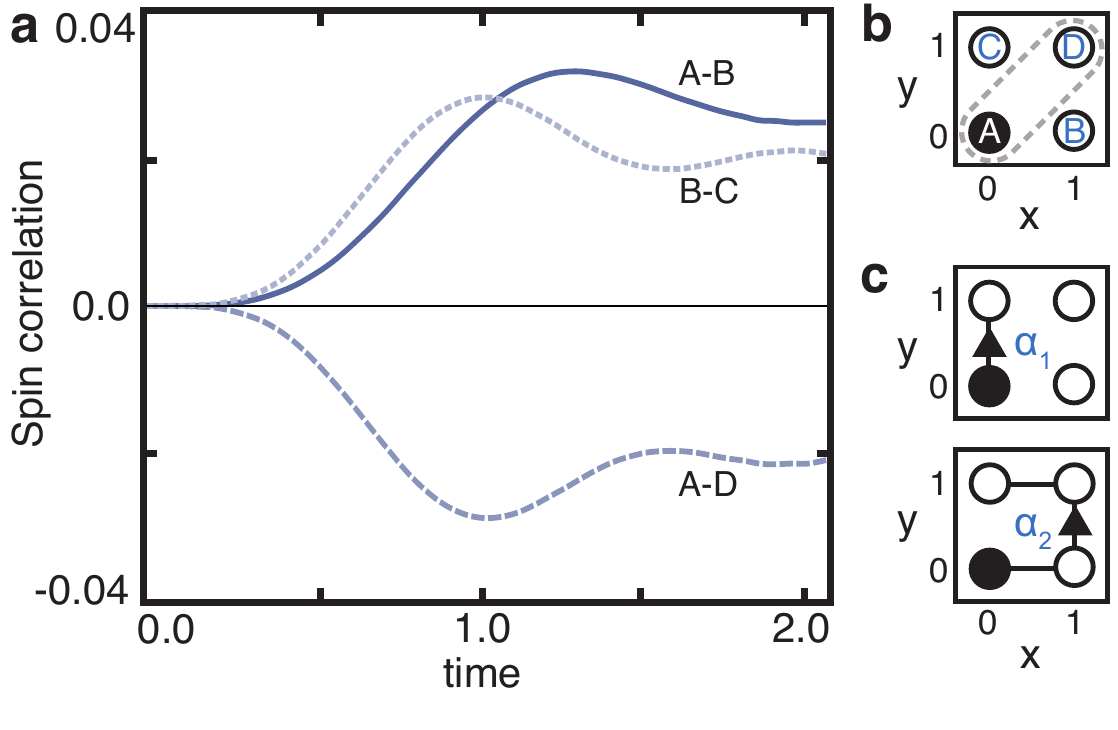}
\caption{{\bf Time-dependent spin correlations.} {\bf a,} Correlations between sites neighboring the origin appear gradually and they stay finite during the time-scale of our calculation. Results are shown for a spin $S=1/2$ system.  Letters $A, B, C$ and $D$ in  {\bf b} denote sites $(0,0), (0,1), (1,0)$ and $(1,1)$, respectively.  Different curves in {\bf a} show correlations between $A$-$B$ (full line), $A$-$D$ (dashed line) and $B$-$C$ (dotted line). 
{\bf c} Lowest order contributions to $A-D$ correlations (gray dashed circle in {\bf b}) arise from interference between paths encircling the two-by-two plaquette, with the hole ending up at $B$ or at $C$. Interference between two such paths $\alpha_1$ (top) and $\alpha_2$ requires non-orthogonality of final spin states. Therefore, all three spins on the plaquette need to be identical.
}
\label{fig:time_dependence}
\end{center}
\end{figure}

\noindent {\bf Spin correlations.}
Fig.~\ref{fig:time_dependence}~a shows that the correlations build up gradually at short times and show slightly oscillating behavior at intermediate times. Whereas correlation between the origin and site $(0,1)$ as well as that between sites $(1,0)$ and $(0,1)$ are ferromagnetic, we find antiferromagnetic correlations between the origin and site $(1,1)$. Fig.~\ref{fig:2Dspincorr} demonstrates that correlations exist between other sites that are further away from the origin. These correlations appear gradually as the hole approaches the surrounding spins. Within the time scale of our calculations, the correlations stay finite. Their long time behavior remains an open question, which could be addressed experimentally.

To illustrate how spin correlations with different signs emerge, let us consider the lowest order contribution to the correlation between the origin $o$ and site $(1,1)$. As shown in Fig.~\ref{fig:time_dependence}~b, the sites of the plaquette containing these two sites are labeled by letters $A=o$ to $D=(1,1)$. We thus investigate the spin correlations $C_{AD}(t)$. In this case the hole needs to end up at sites $j=B$ or $C$ at time t to obtain non-vanishing correlations. Due to symmetry, we only need to consider the case when the hole ends up at site $C$.  As we mentioned earlier, spin correlations can only arise from inequivalent pairs of paths. The lowest order such pair ending at site $C$ is shown in the upper and lower panels of Fig~\ref{fig:time_dependence}~c, and we denote them as $\alpha_1$ and $\alpha_2$, respectively. The diagonal matrix elements of the spin correlator vanish after taking the thermal average over all initial spin configurations, $\langle \hat{\pi}_{\alpha_1}^\dagger \hat{S}_A^z \, \hat{S}_D^z \hat{\pi}_{\alpha_1}\rangle_0 = \langle \hat{S}_B^z \rangle_0 \; \langle \hat{S}_C^z \rangle_0 = 0$, and similarly for path $\alpha_2$. The interference terms between paths $\alpha_1$ and $\alpha_2$, however, yield a non-vanishing contribution for special initial spin states, where $\alpha_1$ and $\alpha_2$ result in the same final spin configuration. As the combined effect of the two paths $\hat{\pi}_{\alpha_2}^\dagger \hat{\pi}_{\alpha_1}$ moves all spins to a neighboring site, the interference term is zero unless all three spins are ferromagnetically aligned. In the infinite temperature system, all spins take random values with equal probability $1/\mathcal{N}$. The probability of all three spins taking on the same configuration is given by $\langle \hat{\pi}_{\alpha_2}^\dagger \hat{\pi}_{\alpha_1} \rangle_0 = 1/\mathcal{N}^2$. As interference terms between identical paths average out to zero due to thermal fluctuation, this term determines the sign of  correlations $C_{AD}(t) = 2 \, (i t) \frac{(-i t)^3}{3!}  \langle \hat{\pi}_{\alpha_2}^\dagger \hat{\pi}_{\alpha_1} \rangle_0 \propto - t^4/(3 \, \mathcal{N}^2)$ at lowest order, which is negative due to the phase factors acquired by the hole along the two paths. Fig~\ref{fig:time_dependence}~a shows that the corresponding correlator $C_{AD}(t)$ stays antiferromagnetic at intermediate times as well, whereas $C_{AB}(t)$ and $C_{BC}(t)$ are ferromagnetic.

In order to evaluate spin correlations and transition probabilities up to any order, we consider two arbitrary paths $\alpha$ and $\beta$ and evaluate their interference $\langle \hat{\pi}_\beta^\dagger \, \hat{\pi}_\alpha \rangle_0$. Permutations created by longer paths can be more complicated than the one shown in Fig.~\ref{fig:time_dependence}~c on the two-by-two plaquette. Since longer paths may intersect each other and themselves, the hole may permute different regions of the lattice independently, as we illustrate in Fig.~\ref{fig:interference}~d. In each of these regions, the spins need to be ferromagnetically aligned to ensure that the initial and the final spin state are not orthogonal. However, the individual regions may take on different ferromagnetic states. These regions can be identified by the separate permutation cycles $\mathcal{C}_a$ of the combined permutation $\hat{\pi}_\beta^\dagger \, \hat{\pi}_\alpha = \Pi_a \mathcal{C}_a$~\cite{Jones1998GroupTheoryBook}. The interference term $\langle \hat{\pi}_\beta^\dagger \hat{\pi}_\alpha \rangle_0$ is thus determined by the probability of the spins being ferromagnetically aligned in each cycle,
\beq
\langle \hat{\pi}_\beta^\dagger \hat{\pi}_\alpha\rangle_0 = \prod_a \frac{1}{\mathcal{N}^{ |\mathcal{C}_a| - 1}}.
\label{eq:overlap}
\eeq
Here, $|\mathcal{C}_a|$ denotes the number of spins in cycle $a$ and $\mathcal{N} = 2S + 1$ is the number of spin degrees of freedom. This interference term also contributes to the spin correlations between sites within the same ferromagnetic region, as we show in Methods. When the spins on sites $\pi_\alpha^\dagger(j)$ and $\pi_\alpha^\dagger(l)$ are within the same ferromagnetic domain, the matrix element $\langle \hat{\pi}^\dagger_\beta S^z_j S^z_l \hat{\pi}_\alpha \rangle$ is simply given by Eq.~\eqref{eq:overlap}. In contrast, it vanishes for all other combination of sites, as the spin correlations between independent domains average out to zero. We determine both the transition probabilities and spin correlations by Monte Carlo sampling the paths and using Eq.~\eqref{eq:overlap} to calculate the interference between each pair of paths (Methods).

When the number of spin degrees of freedom is large, it is very unlikely to find locally ferromagnetic regions in an infinite temperature bath. Since interference of inequivalent paths relies on these domains, these contributions vanish in the limit of large spins $S = \infty$, as Eq.~\eqref{eq:overlap} shows. The strongest spin correlations can be observed in a spin $S=1/2$ system. Furthermore, the interference term is also exponentially suppressed if the paths permute a large number of spins differently. The largest contribution to spin correlations thus arises from the paths that have almost identical effect on the spins. This explains why the induced spin correlations are localized within a few sites in Fig.~\ref{fig:2Dspincorr}.  \\

{\bf Hole dynamics. }
The dependence of the hole's propagation on the spin $S$ of the environment has been demonstrated numerically in Ref.~\onlinecite{Carlstrom2016HolePropagation}. Here, we show that this effect can be attributed to interference between inequivalent paths, that are also responsible for the spin correlations in the environment. Furthermore, we give a simple analytic approximation of the hole's dynamics in a large spin $S=\infty$ environment.

\begin{figure*}[th]
\begin{center}
\includegraphics[width = 16cm]{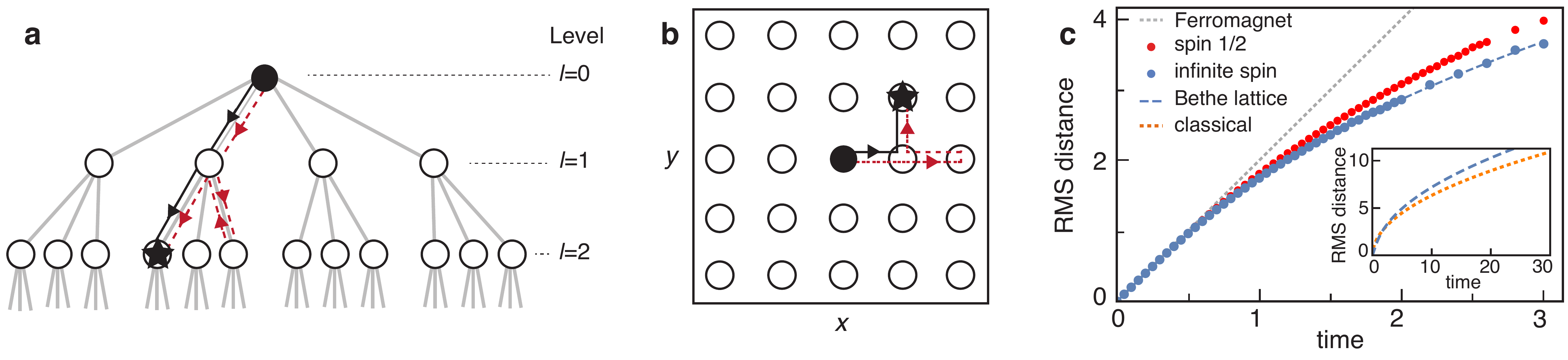}
\caption{{\bf Bethe lattice model.} {\bf a} Bethe lattice of coordination number $z=4$ with the hole (black) at the origin. Solid and dashed lines indicate two interfering paths and the black star denotes their endpoint. These paths correspond to random walks of the hole on the two-dimensional lattice shown in {\bf b}, with the lattice sites denoted by circles. {\bf c} Comparison of the RMS distance of the hole in different models: blue dashed line corresponds to propagation on the Bethe lattice. The gray dashed line, red and blue dots show results in a ferromagnetic ($S=0$), a spin $S=1/2$ and $S=\infty$ degenerate environment, respectively, which are reproduced from Ref.~\onlinecite{Carlstrom2016HolePropagation}. Inset: RMS distance of the hole on the Bethe lattice (blue dashed) and for a classical random walk (orange dots) at long times.}
\label{fig:RMS}
\end{center}
\end{figure*}

In the simplest case of a ferromagnet ($S=0$), all pairs of paths interfere with a maximal amplitude  $\langle \hat{\pi}^\dagger \hat{\pi} \rangle_0 = 1$. As shown in Fig.~\ref{fig:RMS}, the resulting propagation is ballistic, and the root mean squared (RMS) distance of the hole $d_{\rm RMS} = ( \sum_j p_j \, r_j^2)^{1/2}$ grows linearly in time, with $r_j$ denoting the distance of site $j$ from the origin. 
However, the propagation of the hole is slowed down in environments of finite spin, as a result of the suppression of interference terms between inequivalent paths, shown in Eq.~\eqref{eq:overlap}. Thus, in the $S \to \infty$ limit, only equivalent paths contribute to the dynamics. 

In order to gain insight into this limit, we investigate the propagation of the hole on the Bethe lattice~\cite{Ostilli2012CayleyTreesAndBetheLattices}, shown in Fig.~\ref{fig:RMS}~a. The Bethe lattice is a tree graph, with the origin at the root level $l=0$. Each site has $z=4$ neighbors, that can be identified with left, right, up and down steps on the two-dimensional lattice. Each random walk on the Bethe lattice can thus be identified with one on the square lattice. The position of the hole on the Bethe lattice keeps full information of its two-dimensional path up to self-retracing components, and two paths interfere if and only if their endpoints are the same. Due to the geometrical constraint imposed by the graph, interfering paths cannot include loops. In particular, the Bethe lattice only allows interference between equivalent paths, that are identical except for self-retracing components, see Fig.~\ref{fig:RMS}~b. This construction covers most of the phase space of equivalent pairs, which determine the hole's propagation in the $S=\infty$ environment. Therefore, the hole's dynamics in this system is expected to be well approximated by the Bethe lattice construction.

As two interfering paths on the Bethe lattice always permute the spins the same way, the hole's dynamics on the Bethe lattice decouples completely from that of the spins. It therefore becomes a single particle problem that can be solved analytically (Methods). This behavior is reminiscent of the physics of spin-charge separation in a one-dimensional lattice ~\cite{Ogata1990SpinChargeSeparation1DBetheAnsatz, Kruis2004SpinChargeSeparationInLuttingerLiquids, Hilker2017HiddenAFM}, which is equivalent to the Bethe lattice of coordination number $z=2$. In that case a hole moves coherently in the lattice, while keeping the order of the spins unchanged. Although the spin-configuration depends on the hole position, this does not introduce correlations between the spins. In two dimensions, the dynamics on the $z=4$ Bethe lattice is more subtle. Due to interference between equivalent paths, the average level of the graph grows linearly in time similar to one dimensional systems  (Supplementary Material). However, this does not manifest as ballistic propagation on the square lattice. The RMS distance of sites on level $l$ of the Bethe lattice becomes $d_l = (2 l - \frac{3}{2}(1-3^{-l}))^{1/2}$, which grows as $d_l \sim \sqrt{2l}$ at large distances (Methods). Therefore, instead of ballistic propagation, we find that the hole shows diffusive behavior at long times $d_{\rm RMS} \sim \sqrt{2 D_{\rm Bethe} t}$, with a diffusion constant $D_{\rm Bethe} \approx 2.73$. However, the diffusion is faster than that of a classical random walk, having a diffusion constant $D_{\rm cl} = 2$ (Methods).

The RMS distance is shown for different models in Fig.~\ref{fig:RMS}. We find the usual ballistic propagation in the $S=0$ ferromagnet, whereas the hole appears to cross over from ballistic to diffusive behavior in the $S>0$ case at intermediate times, as was discussed in Ref.~\onlinecite{Carlstrom2016HolePropagation}. Fig.~\ref{fig:RMS} shows, that although the interference between inequivalent paths are small, they lead to faster propagation in the $S=1/2$ spin environment than in the infinite spin case. This difference is therefore due to the same interference terms that give rise to the spin correlations in the environment. The insight of our work is that the RMS distance of the $S=\infty$ model and the Bethe lattice agree within error bars of our simulation, indicating that the behavior of the two models are in very good agreement at short and intermediate times. Therefore we expect that, similar to the Bethe lattice propagation, the hole's dynamics will cross over from ballistic to diffusive behavior in an infinite spin environment.  \\

\noindent {\bf Discussion.}
The spin correlations presented in this paper demonstrate a general paradigm of how originally completely disordered environment can acquire correlations due to quantum interference in the course of non-equilibrium dynamics of a particle. We emphasize that this mechanism is fundamentally different from the interaction-induced correlations in the bath discussed in Refs.~\cite{Landau1933Polaron, Frohlich1954Polaron, Zhang1991tJHolePropagation, Mishchenko2000DiagrammaticQMC,  Mishchenko2001tJModel, White2001tJPolaronDMRG, Shchadilova2016QuantumDynamicsOfUltracoldBosePolarons, Jorgensen2016ObservationOfAttractiveAndRepulsivePolarons, Hu2016BosePolaronsInTheStronglyInteractingRegime}
and can be observed at infinite temperature of the spin bath. Experimental realization of this phenomenon using ultracold atoms would provide an ideal opportunity for the study of entanglement between a particle and its environment that is usually challenging in other setups due to the fast decoherence in the environment. These experiments could also provide information on the long time dynamics of spin correlation, that remains an open question.

Further interesting questions arise about the effect of spin interactions, which appear naturally in experiments with smaller on-site interactions, whether in ultracold atomic or electronic Mott insulators. This would affect spin correlations even in an infinite temperature spin environment with interactions, as the spin correlations can be decohered by magnetic excitations of the environment.  At lower temperatures, the energy cost of permuting spins leads to a strongly renormalized dynamics of the hole~\cite{Bulaevski1968, BrinkmanRice1970, Trugman1988, SchmittRinkVarma1988SpectralFunctionOfTheHoleInAQuantumAFM, Kane1989MotionOfASingleHoleInAQuantumAFM,  Zhang1991tJHolePropagation, Mishchenko2001tJModel, White2001tJPolaronDMRG, Mierzejewski2011NonEqQDynamics, Bonca2012HolePairDynamics, DalConte2015RetardedInteractionInCuprates}. 
Understanding this limit, and especially the interplay of multiple holes with the environment could also lead to a better understanding of the role of doping in the cuprate phase diagram~\cite{Bulaevski1968, Shastry1990NagaokaInsability, WenLee2006HighTcReview, Sachdev2007Book, Auerbach2012Book, Bonca2012HolePairDynamics, DalConte2015RetardedInteractionInCuprates}. 
\\

\noindent {\bf Acknowledgements} \\
We thank J. Carlstr\"om, N. Prokof'ev, G. Zar\'and and J. van den Brink for enlightening discussions. \marton{The authors acknowledge support from Harvard-MIT CUA, NSF Grant No. DMR-1308435, AFOSR Quantum Simulation MURI, AFOSR grant number FA9550-16-1-0323, the Moore Foundation, the Harvard Quantum Optics Center and the Swiss National Foundation. I. L. was supported by the Hungarian research grant OTKA Nos. K105149 and SNN118028.} \\

\noindent {\bf Author contributions} \\
M. K.-N. and I. L. performed numerical and analytical calculations and wrote the paper. All authors developed theoretical concepts, discussed the results and commented on the manuscript. \\

\noindent {\bf Additional information} \\
{\bf Competing financial interests: } The authors declare no competing financial interest.

\newpage

\appendix

 \begin{center}
    {\bf METHODS}
  \end{center}

\section{Effect of spin conservation}
Since the hole cannot create spin flips, the total spin $S_{tot}^z = \sum_j S_j^z$ is conserved. The time dependence of the square of this operator can be expressed in terms of spin correlations, $\frac{1}{S^2} \,\langle c_o^\dagger \, e^{i H t} \, (S_{tot}^z)^2 \, e^{- i H t} \, c_o^\dagger \rangle = \sum_j  C_{j j}(t) + \sum_{j \ne l}  C_{j l}(t)$. The diagonal spin correlations are simply given by $C_{jj}(t) = 1 - p_j(t)$, the sum of these correlations is thus constant $\sum_j  C_{j j}(t) = M - 1$, where $M$ is the number of sites of the system. We thus find, that the sum of off-diagonal spin correlations is conserved. Since it is zero in the initial state, we arrive at the sum rule $\sum_{j \ne l} C_{jl}(t) = 0$.

\section{Interference contribution to spin correlations}
The contribution of paths $\alpha$ and $\beta$ to the spin correlations $C_{jl}(t)$ is given by $\frac{(-i t)^{n_\beta} \, (i t)^{n_\alpha}}{n_\beta! \, n_\alpha!}\, \langle \hat{\pi}^\dagger_\beta S^z_j S^z_l \hat{\pi}_\alpha \rangle $. Here, $n_\beta$ and $n_\alpha$ denote the lengths of these paths. The non-orthogonality of the initial and the final spin states requires each permutation cycle of the combined path $\hat{\pi}_\beta^\dagger \, \hat{\pi}_\alpha$ to be ferromagnetic. Therefore, if the spins on sites $j^\prime = \pi_\alpha^\dagger(j)$ and $l^\prime = \pi_\alpha^\dagger(l)$ are in the same permutation cycle,  the above expectation value becomes $\langle \, \hat{\pi}^\dagger_\beta \, S^z_j S^z_l \, \hat{\pi}_\alpha \rangle = \langle \, \hat{\pi}^\dagger_\beta \hat{\pi}_\alpha \, S^z_{j^\prime} \, S^z_{l^\prime} \rangle = \frac{S(S+1)}{3} \langle \, \hat{\pi}^\dagger_\beta \hat{\pi}_\alpha \rangle$. The $S$ dependent prefactor in the previous equation arises from averaging the spin operators over all $(2S+1)$ possible ferromagnetic spin configurations. For all other pairs of sites, the spins are independent, and the expectation value $\langle \, \hat{\pi}^\dagger_\beta S^z_j S^z_l \hat{\pi}_\alpha \rangle$ thus averages out to zero in the infinite temperature spin environment.

\section{Real-time quantum Monte Carlo algorithm}
We sample the time evolution operator using stochastic series expansion quantum Monte Carlo~\cite{Sandvik1991StochasticSeriesExpansion, Carlstrom2016HolePropagation}. At the beginning of the simulation, we generate of the order of $2 \times 10^8$ random walk paths. The permutations generated by these paths are binned, and their phase factors $i^n$ are added. The $\frac{t^n}{n!}$ amplitudes are taken into account by sampling the path lengths $n$ according to a Posson distribution $\mathbb{P}_n \propto \frac{(z t)^n}{n!}$. The resulting amplitudes $\lambda$ are stored together with the corresponding permutations $\hat{\pi}$ as pairs $(\lambda, \hat{\pi})$. We take all possible combinations of forward $(\lambda_\alpha, \hat{\pi}_\alpha)$ and backward $(\lambda_\beta, \hat{\pi}_\beta)$ time evolution bins. We evaluate the many-body trace associated with each pair using Eq.~\eqref{eq:overlap} exactly. We add the interference contribution $\lambda_\beta^* \, \lambda_\alpha \langle\hat{\pi}_\beta^\dagger \hat{\pi}_\alpha \rangle$ to the histogram of the transition probabilities $\tilde{p}_j(t)$ and that of spin correlations $\tilde{C}_{jl}(t)$ for each site $j$ and $l$. At the end of the simulation, we normalize the histograms $\tilde{p}_j(t)$ and $\tilde{C}_{jl}(t)$ by dividing them by $\sum_j \tilde{p}_j(t)$.

Evaluating the infinte temperature spin averages in Eq.~\eqref{eq:overlap} exactly allows us to sample the spin correlation with very small error bars, as compared to performing numerical averaging over different spin configurations~\onlinecite{Carlstrom2016HolePropagation}. However, around time $t\sim 1.8$, the phase space of important paths becomes significantly larger than the number of our samples.  Since the number of path bins $L$ also becomes very large, calculating the interference contributions of all the $L\times L$ pairs of path bins becomes impractical. Therefore, at longer times $2 < t < 3$, we calculate the RMS distance using a slightly modified version of the algorithm of Ref.~\cite{Carlstrom2016HolePropagation}, but with different set of forward and backward paths. Although this method provides noisy spin correlation data, it determines RMS distance at longer times very accurately, and requires only of the order of $L$ steps.

\section{Sampling of forward and backward time evolution paths}
In contrast to Ref.~\cite{Carlstrom2016HolePropagation}, we sample the forward $e^{-i H t}$ and backward $e^{i H t}$ time evolution paths independently. The independent sampling becomes important at times longer than $t \sim 1.8$, when the phase space of paths becomes so large, that the quantum Monte Carlo procedure can sample it only sparsely. At these long times, two typical paths $\alpha$ and $\beta$ will in general be long, and they therefore enclose large loops. According to Eq.~\eqref{eq:overlap}, their interference $\langle \hat{\pi}_\beta^\dagger \, \hat{\pi}_\alpha \rangle_0$ is exponentially small. 

Choosing the forward and backward time evolution paths from the same sample would lead to large systematic errors. In contrast to independent sampling, this procedure would over-sample those cases when the forward and backward paths are identical. However, these pairs have an interference of $\langle \hat{\pi}_\alpha^\dagger \, \hat{\pi}_\alpha \rangle_0 = 1$, in contrast to the typically  exponentially small interference of non-identical pairs. Therefore, the pairs consisting of identical paths overwhelm contributions from non-identical paths, leading to incorrect results. By sampling the forward and backward evolution paths independently, these errors can be avoided.

\section{Propagation on the Bethe lattice}
The hole's propagation on the Bethe lattice can be solved analytically, as we show in the Supplementary Material. Here, we present a shorter, recursive solution. Expanding the time evolution in terms of random walks, we find that the wave function of the hole at level $l$ is given by $\psi_{l}(t) = \frac{1}{M_l}  \sum_{n = 0}^\infty \frac{(-i z t)^n}{n!} \rho_{n, l},$ where $M_l$ denotes the number of sites level $l$, with $M_l = 1/(z \, (z-1)^{l-1})$ for $l \geq 1$, and $M_0 = 1$. The matrix $\rho_{n,l}$ denotes the probability that a random walk path of length $n$ ends up at level $l$. These probabilities can be determined using simple recurrence relations. At all levels $l \geq 1$, the probability of taking a step one level down on the graph is $3/4$ and taking a step up has a probability $1/4$. At the origin, the walker goes to level $l=1$ with probability $1$. This leads to the following recurrence relations $\rho_{n+1, \, l} = \frac{3}{4} \, \rho_{n, \, l-1} + \frac{1}{4} \, \rho_{n, \, l+1}$ for $l \geq 2$, and for levels $l = 0, 1$ we get  $\rho_{n+1, \, 1} = \rho_{n, \, 0} + \frac{1}{4} \rho_{n, \, 2}$ and $\rho_{n,\, 0} = \frac{1}{4} \rho_{n,\, 1}$. We solve these equations iteratively, starting from the initial condition $\rho_{0,l} = \delta_{0,l}$.

\section{RMS distance of sites on the Bethe lattice} \label{app:1}
We determine the RMS distance $d_l$ of sites on level $l$, using an iterative procedure. When mapping the sites at level $l$ of the Bethe lattice to the square lattice, we get the end points of all possible random walks of length $l$, involving no self-retracing components. In order to calculate the RMS distance for the end points of such random walks, we write down a recurrence relation between $d_l^2$ and $d_{l-1}^2$. Let $(x_l, y_l)$ denote the hole's displacement in its $l$th step. We can assume without the loss of generality that the first step was taken to the right. The RMS distance of the endpoint can be written as $d_l^2 =d_{l-1}^2 + 2 \overline{x}_{l-1} + 1$, where $\overline{x}_{l-1}$ denotes the average number of right steps in the remaining path. This quantity is non-zero since the left-right symmetry of the walk is broken due to the initial step. However, after the first time the hole moves in the up or down direction, the left-right symmetry of the model is restored, and the remaining part of the path does not contribute to  $\overline{x}_{l-1}$. The probability of taking $n$ steps to the right, and then an up or down step is given by $(1/3)^n \, (2/3)$. Summing up the series for all $n < l-1$, and adding the probability of taking all remaining $l-1$ steps to the right, $(1/3)^{l-1}$, leads to $\overline{x}_{l-1} = (1-3^{-(l-1)})/2$. We thus obtain the recurrence relation $d_l^2 = d_{l-1}^2 + 2 - 3^{-(l-1)}$, which can be solved exactly, yielding $d_l^2 = 2 l - \frac{3}{2} (1- 3^{-l})$.

\section{Comparison with classical random walks}
We compare the quantum dynamics of the hole to that of a classical particle performing Brownian motion. With the particle starting from the origin, its time evolution is governed by a transition rate matrix, assigning the transition rate $1$ to each of its neighboring sites $i$ and $j$. The probability distribution of the particle thus follows a classical diffusion equation, with a diffusion constant $D_{\rm cl}=2$.

\bibliography{inf_temp_spins_refs}

\begin{thebibliography}{10}
\expandafter\ifx\csname url\endcsname\relax
  \def\url#1{\texttt{#1}}\fi
\expandafter\ifx\csname urlprefix\endcsname\relax\def\urlprefix{URL }\fi
\providecommand{\bibinfo}[2]{#2}
\providecommand{\eprint}[2][]{\url{#2}}

\bibitem{Friedman2000QuantumInterference}
\bibinfo{author}{Friedman, J.~R.}, \bibinfo{author}{Patel, V.},
  \bibinfo{author}{Chen, W.}, \bibinfo{author}{Tolpygo, S.} \&
  \bibinfo{author}{Lukens, J.~E.}
\newblock \bibinfo{title}{Quantum superposition of distinct macroscopic
  states}.
\newblock \emph{\bibinfo{journal}{nature}} \textbf{\bibinfo{volume}{406}},
  \bibinfo{pages}{43--46} (\bibinfo{year}{2000}).

\bibitem{Hofstetter2009CooperPairSplitting}
\bibinfo{author}{Hofstetter, L.}, \bibinfo{author}{Csonka, S.},
  \bibinfo{author}{Nyg{\aa}rd, J.} \& \bibinfo{author}{Sch{\"o}nenberger, C.}
\newblock \bibinfo{title}{Cooper pair splitter realized in a two-quantum-dot
  y-junction}.
\newblock \emph{\bibinfo{journal}{Nature}} \textbf{\bibinfo{volume}{461}},
  \bibinfo{pages}{960--963} (\bibinfo{year}{2009}).

\bibitem{Andrews1997InterferenceOfBECs}
\bibinfo{author}{Andrews, M.} \emph{et~al.}
\newblock \bibinfo{title}{Observation of interference between two bose
  condensates}.
\newblock \emph{\bibinfo{journal}{Science}} \textbf{\bibinfo{volume}{275}},
  \bibinfo{pages}{637--641} (\bibinfo{year}{1997}).

\bibitem{Simmonds2001QuantumInterferenceOfHe3}
\bibinfo{author}{Simmonds, R.}, \bibinfo{author}{Marchenkov, A.},
  \bibinfo{author}{Hoskinson, E.}, \bibinfo{author}{Davis, J.} \&
  \bibinfo{author}{Packard, R.}
\newblock \bibinfo{title}{Quantum interference of superfluid 3he}.
\newblock \emph{\bibinfo{journal}{Nature}} \textbf{\bibinfo{volume}{412}},
  \bibinfo{pages}{55--58} (\bibinfo{year}{2001}).

\bibitem{Chin2006SuperfluidInterference}
\bibinfo{author}{Chin, J.~K.} \emph{et~al.}
\newblock \bibinfo{title}{Evidence for superfluidity of ultracold fermions in
  an optical lattice}.
\newblock \emph{\bibinfo{journal}{Nature}} \textbf{\bibinfo{volume}{443}},
  \bibinfo{pages}{961--964} (\bibinfo{year}{2006}).

\bibitem{Hubbard1970SpinDiffusion}
\bibinfo{author}{Blume, M.} \& \bibinfo{author}{Hubbard, J.}
\newblock \bibinfo{title}{Spin correlation functions at high temperatures}.
\newblock \emph{\bibinfo{journal}{Phys. Rev. B}} \textbf{\bibinfo{volume}{1}},
  \bibinfo{pages}{3815--3830} (\bibinfo{year}{1970}).
\newblock \urlprefix\url{http://link.aps.org/doi/10.1103/PhysRevB.1.3815}.

\bibitem{ChertkovKolokov1994SpinDiffusion}
\bibinfo{author}{Chertkov, M.} \& \bibinfo{author}{Kolokolov, I.}
\newblock \bibinfo{title}{Long-time dynamics of the infinite-temperature
  heisenberg magnet}.
\newblock \emph{\bibinfo{journal}{Phys. Rev. B}} \textbf{\bibinfo{volume}{49}},
  \bibinfo{pages}{3592--3595} (\bibinfo{year}{1994}).
\newblock \urlprefix\url{http://link.aps.org/doi/10.1103/PhysRevB.49.3592}.

\bibitem{Lovesey1994SpinDiffusion}
\bibinfo{author}{Lovesey, S.}, \bibinfo{author}{Engdahl, E.},
  \bibinfo{author}{Cuccoli, A.}, \bibinfo{author}{Tognetti, V.} \&
  \bibinfo{author}{Balcar, E.}
\newblock \bibinfo{title}{Time-dependent spin correlations in the heisenberg
  magnet at infinite temperature}.
\newblock \emph{\bibinfo{journal}{Journal of Physics: Condensed Matter}}
  \textbf{\bibinfo{volume}{6}}, \bibinfo{pages}{L521} (\bibinfo{year}{1994}).

\bibitem{Engel2007Photosynthesis}
\bibinfo{author}{Engel, G.~S.} \emph{et~al.}
\newblock \bibinfo{title}{Evidence for wavelike energy transfer through quantum
  coherence in photosynthetic systems}.
\newblock \emph{\bibinfo{journal}{Nature}} \textbf{\bibinfo{volume}{446}},
  \bibinfo{pages}{782--786} (\bibinfo{year}{2007}).

\bibitem{Collini2010Photosynthesis}
\bibinfo{author}{Collini, E.} \emph{et~al.}
\newblock \bibinfo{title}{Coherently wired light-harvesting in photosynthetic
  marine algae at ambient temperature}.
\newblock \emph{\bibinfo{journal}{Nature}} \textbf{\bibinfo{volume}{463}},
  \bibinfo{pages}{644--647} (\bibinfo{year}{2010}).

\bibitem{Panitchayangkoon2010Photosynthesis}
\bibinfo{author}{Panitchayangkoon, G.} \emph{et~al.}
\newblock \bibinfo{title}{Long-lived quantum coherence in photosynthetic
  complexes at physiological temperature}.
\newblock \emph{\bibinfo{journal}{Proceedings of the National Academy of
  Sciences}} \textbf{\bibinfo{volume}{107}}, \bibinfo{pages}{12766--12770}
  (\bibinfo{year}{2010}).

\bibitem{Ritz2000BirdNavigation}
\bibinfo{author}{Ritz, T.}, \bibinfo{author}{Adem, S.} \&
  \bibinfo{author}{Schulten, K.}
\newblock \bibinfo{title}{A model for photoreceptor-based magnetoreception in
  birds}.
\newblock \emph{\bibinfo{journal}{Biophysical journal}}
  \textbf{\bibinfo{volume}{78}}, \bibinfo{pages}{707--718}
  (\bibinfo{year}{2000}).

\bibitem{Turin1996Smelling}
\bibinfo{author}{Turin, L.}
\newblock \bibinfo{title}{A spectroscopic mechanism for primary olfactory
  reception}.
\newblock \emph{\bibinfo{journal}{Chemical Senses}}
  \textbf{\bibinfo{volume}{21}}, \bibinfo{pages}{773--791}
  (\bibinfo{year}{1996}).

\bibitem{NielsenChuang2002QuantumInfoBook}
\bibinfo{author}{Nielsen, M.~A.} \& \bibinfo{author}{Chuang, I.}
\newblock \bibinfo{title}{Quantum computation and quantum information}
  (\bibinfo{year}{2002}).

\bibitem{Prokofev2006DecoherenceAndQuantumWalks}
\bibinfo{author}{Prokof'ev, N.~V.} \& \bibinfo{author}{Stamp, P. C.~E.}
\newblock \bibinfo{title}{Decoherence and quantum walks: Anomalous diffusion
  and ballistic tails}.
\newblock \emph{\bibinfo{journal}{Phys. Rev. A}} \textbf{\bibinfo{volume}{74}},
  \bibinfo{pages}{020102} (\bibinfo{year}{2006}).
\newblock \urlprefix\url{http://link.aps.org/doi/10.1103/PhysRevA.74.020102}.

\bibitem{Morello2006DecoherenceInSpinQubitNetworks}
\bibinfo{author}{Morello, A.}, \bibinfo{author}{Stamp, P. C.~E.} \&
  \bibinfo{author}{Tupitsyn, I.~S.}
\newblock \bibinfo{title}{Pairwise decoherence in coupled spin qubit networks}.
\newblock \emph{\bibinfo{journal}{Phys. Rev. Lett.}}
  \textbf{\bibinfo{volume}{97}}, \bibinfo{pages}{207206}
  (\bibinfo{year}{2006}).
\newblock
  \urlprefix\url{http://link.aps.org/doi/10.1103/PhysRevLett.97.207206}.

\bibitem{Novoselov2007RoomTemperatureQuantumHall}
\bibinfo{author}{Novoselov, K.~S.} \emph{et~al.}
\newblock \bibinfo{title}{Room-temperature quantum hall effect in graphene}.
\newblock \emph{\bibinfo{journal}{Science}} \textbf{\bibinfo{volume}{315}},
  \bibinfo{pages}{1379--1379} (\bibinfo{year}{2007}).

\bibitem{SethLloyd2011QuantumCoherenceInBiologicalSystems}
\bibinfo{author}{Lloyd, S.}
\newblock \bibinfo{title}{Quantum coherence in biological systems}.
\newblock In \emph{\bibinfo{booktitle}{Journal of Physics: Conference Series}},
  vol. \bibinfo{volume}{302}, \bibinfo{pages}{012037}
  (\bibinfo{organization}{IOP Publishing}, \bibinfo{year}{2011}).

\bibitem{Zurek1991Decoherence}
\bibinfo{author}{Zurek, W.~H.}
\newblock \bibinfo{title}{Decoherence and the transition from quantum to
  classical}.
\newblock \emph{\bibinfo{journal}{Physics today}}
  \textbf{\bibinfo{volume}{44}}, \bibinfo{pages}{36--44}
  (\bibinfo{year}{1991}).

\bibitem{Jordens2008FermionicMI}
\bibinfo{author}{J{\"o}rdens, R.}, \bibinfo{author}{Strohmaier, N.},
  \bibinfo{author}{G{\"u}nter, K.}, \bibinfo{author}{Moritz, H.} \&
  \bibinfo{author}{Esslinger, T.}
\newblock \bibinfo{title}{A mott insulator of fermionic atoms in an optical
  lattice}.
\newblock \emph{\bibinfo{journal}{Nature}} \textbf{\bibinfo{volume}{455}},
  \bibinfo{pages}{204--207} (\bibinfo{year}{2008}).

\bibitem{BlochSUN}
\bibinfo{author}{Hofrichter, C.} \emph{et~al.}
\newblock \bibinfo{title}{Direct probing of the mott crossover in the su (n)
  fermi-hubbard model}.
\newblock \emph{\bibinfo{journal}{Physical Review X}}
  \textbf{\bibinfo{volume}{6}}, \bibinfo{pages}{021030} (\bibinfo{year}{2016}).

\bibitem{Nagaoka1965Ferromagnetism}
\bibinfo{author}{Nagaoka, Y.}
\newblock \bibinfo{title}{Ground state of correlated electrons in a narrow
  almost half-filled s band}.
\newblock \emph{\bibinfo{journal}{Solid State Communications}}
  \textbf{\bibinfo{volume}{3}}, \bibinfo{pages}{409 -- 412}
  (\bibinfo{year}{1965}).
\newblock
  \urlprefix\url{http://www.sciencedirect.com/science/article/pii/0038109865902668}.

\bibitem{Nagaoka1966Ferromagnetism}
\bibinfo{author}{Nagaoka, Y.}
\newblock \bibinfo{title}{Ferromagnetism in a narrow, almost half-filled $s$
  band}.
\newblock \emph{\bibinfo{journal}{Phys. Rev.}} \textbf{\bibinfo{volume}{147}},
  \bibinfo{pages}{392--405} (\bibinfo{year}{1966}).
\newblock \urlprefix\url{http://link.aps.org/doi/10.1103/PhysRev.147.392}.

\bibitem{Auerbach2012Book}
\bibinfo{author}{Auerbach, A.}
\newblock \emph{\bibinfo{title}{Interacting electrons and quantum magnetism}}
  (\bibinfo{publisher}{Springer Science \& Business Media},
  \bibinfo{year}{2012}).

\bibitem{Carlstrom2016HolePropagation}
\bibinfo{author}{Carlstr\"om, J.}, \bibinfo{author}{Prokof'ev, N.} \&
  \bibinfo{author}{Svistunov, B.}
\newblock \bibinfo{title}{Quantum walk in degenerate spin environments}.
\newblock \emph{\bibinfo{journal}{Phys. Rev. Lett.}}
  \textbf{\bibinfo{volume}{116}}, \bibinfo{pages}{247202}
  (\bibinfo{year}{2016}).
\newblock
  \urlprefix\url{http://link.aps.org/doi/10.1103/PhysRevLett.116.247202}.

\bibitem{Landau1933Polaron}
\bibinfo{author}{Landau, L.}
\newblock \bibinfo{title}{{\"U}ber die bewegung der elektronen in
  kristalgitter}.
\newblock \emph{\bibinfo{journal}{Phys. Z. Sowjetunion}}
  \textbf{\bibinfo{volume}{3}}, \bibinfo{pages}{644--645}
  (\bibinfo{year}{1933}).

\bibitem{Frohlich1954Polaron}
\bibinfo{author}{Fr{\"o}hlich, H.}
\newblock \bibinfo{title}{Electrons in lattice fields}.
\newblock \emph{\bibinfo{journal}{Advances in Physics}}
  \textbf{\bibinfo{volume}{3}}, \bibinfo{pages}{325--361}
  (\bibinfo{year}{1954}).

\bibitem{Mishchenko2000DiagrammaticQMC}
\bibinfo{author}{Mishchenko, A.}, \bibinfo{author}{Prokof’ev, N.},
  \bibinfo{author}{Sakamoto, A.} \& \bibinfo{author}{Svistunov, B.}
\newblock \bibinfo{title}{Diagrammatic quantum monte carlo study of the
  fr{\"o}hlich polaron}.
\newblock \emph{\bibinfo{journal}{Physical Review B}}
  \textbf{\bibinfo{volume}{62}}, \bibinfo{pages}{6317} (\bibinfo{year}{2000}).

\bibitem{Zhang1991tJHolePropagation}
\bibinfo{author}{Zhang, Q.} \& \bibinfo{author}{Whaley, K.~B.}
\newblock \bibinfo{title}{Exact time-dependent propagation of vacancy motion in
  the t-j limit of the two-dimensional hubbard hamiltonian}.
\newblock \emph{\bibinfo{journal}{Phys. Rev. B}} \textbf{\bibinfo{volume}{43}},
  \bibinfo{pages}{11062--11070} (\bibinfo{year}{1991}).
\newblock \urlprefix\url{http://link.aps.org/doi/10.1103/PhysRevB.43.11062}.

\bibitem{Mishchenko2001tJModel}
\bibinfo{author}{Mishchenko, A.~S.}, \bibinfo{author}{Prokof'ev, N.~V.} \&
  \bibinfo{author}{Svistunov, B.~V.}
\newblock \bibinfo{title}{Single-hole spectral function and spin-charge
  separation in the $t\ensuremath{-}j$ model}.
\newblock \emph{\bibinfo{journal}{Phys. Rev. B}} \textbf{\bibinfo{volume}{64}},
  \bibinfo{pages}{033101} (\bibinfo{year}{2001}).
\newblock \urlprefix\url{http://link.aps.org/doi/10.1103/PhysRevB.64.033101}.

\bibitem{White2001tJPolaronDMRG}
\bibinfo{author}{White, S.~R.} \& \bibinfo{author}{Affleck, I.}
\newblock \bibinfo{title}{Density matrix renormalization group analysis of the
  nagaoka polaron in the two-dimensional $t\ensuremath{-}j$ model}.
\newblock \emph{\bibinfo{journal}{Phys. Rev. B}} \textbf{\bibinfo{volume}{64}},
  \bibinfo{pages}{024411} (\bibinfo{year}{2001}).
\newblock \urlprefix\url{http://link.aps.org/doi/10.1103/PhysRevB.64.024411}.

\bibitem{Shchadilova2016QuantumDynamicsOfUltracoldBosePolarons}
\bibinfo{author}{Shchadilova, Y.~E.}, \bibinfo{author}{Schmidt, R.},
  \bibinfo{author}{Grusdt, F.} \& \bibinfo{author}{Demler, E.}
\newblock \bibinfo{title}{Quantum dynamics of ultracold bose polarons}.
\newblock \emph{\bibinfo{journal}{Physical Review Letters}}
  \textbf{\bibinfo{volume}{117}}, \bibinfo{pages}{113002}
  (\bibinfo{year}{2016}).

\bibitem{Jorgensen2016ObservationOfAttractiveAndRepulsivePolarons}
\bibinfo{author}{J{\o}rgensen, N.~B.} \emph{et~al.}
\newblock \bibinfo{title}{Observation of attractive and repulsive polarons in a
  bose-einstein condensate}.
\newblock \emph{\bibinfo{journal}{Physical Review Letters}}
  \textbf{\bibinfo{volume}{117}}, \bibinfo{pages}{055302}
  (\bibinfo{year}{2016}).

\bibitem{Hu2016BosePolaronsInTheStronglyInteractingRegime}
\bibinfo{author}{Hu, M.-G.} \emph{et~al.}
\newblock \bibinfo{title}{Bose polarons in the strongly interacting regime}.
\newblock \emph{\bibinfo{journal}{Physical Review Letters}}
  \textbf{\bibinfo{volume}{117}}, \bibinfo{pages}{055301}
  (\bibinfo{year}{2016}).

\bibitem{Mozyrsky1998AdiabaticDecoherence}
\bibinfo{author}{Mozyrsky, D.} \& \bibinfo{author}{Privman, V.}
\newblock \bibinfo{title}{Adiabatic decoherence}.
\newblock \emph{\bibinfo{journal}{Journal of statistical Physics}}
  \textbf{\bibinfo{volume}{91}}, \bibinfo{pages}{787--799}
  (\bibinfo{year}{1998}).

\bibitem{Gangopadhyay2001DissipationlessDecoherence}
\bibinfo{author}{Gangopadhyay, G.}, \bibinfo{author}{Kumar, M.~S.} \&
  \bibinfo{author}{Dattagupta, S.}
\newblock \bibinfo{title}{On dissipationless decoherence}.
\newblock \emph{\bibinfo{journal}{Journal of Physics A: Mathematical and
  General}} \textbf{\bibinfo{volume}{34}}, \bibinfo{pages}{5485}
  (\bibinfo{year}{2001}).

\bibitem{Prokof2000TheoryOfSpinBath}
\bibinfo{author}{Prokof'ev, N.} \& \bibinfo{author}{Stamp, P.}
\newblock \bibinfo{title}{Theory of the spin bath}.
\newblock \emph{\bibinfo{journal}{Reports on Progress in Physics}}
  \textbf{\bibinfo{volume}{63}}, \bibinfo{pages}{669} (\bibinfo{year}{2000}).

\bibitem{Unruh2012DecoherenceWithoutDissipation}
\bibinfo{author}{Unruh, W.}
\newblock \bibinfo{title}{Decoherence without dissipation}.
\newblock \emph{\bibinfo{journal}{Philosophical Transactions of the Royal
  Society of London A: Mathematical, Physical and Engineering Sciences}}
  \textbf{\bibinfo{volume}{370}}, \bibinfo{pages}{4454--4459}
  (\bibinfo{year}{2012}).

\bibitem{Ostilli2012CayleyTreesAndBetheLattices}
\bibinfo{author}{Ostilli, M.}
\newblock \bibinfo{title}{Cayley trees and bethe lattices: A concise analysis
  for mathematicians and physicists}.
\newblock \emph{\bibinfo{journal}{Physica A: Statistical Mechanics and its
  Applications}} \textbf{\bibinfo{volume}{391}}, \bibinfo{pages}{3417--3423}
  (\bibinfo{year}{2012}).

\bibitem{WenLee2006HighTcReview}
\bibinfo{author}{Lee, P.~A.}, \bibinfo{author}{Nagaosa, N.} \&
  \bibinfo{author}{Wen, X.-G.}
\newblock \bibinfo{title}{Doping a mott insulator: Physics of high-temperature
  superconductivity}.
\newblock \emph{\bibinfo{journal}{Rev. Mod. Phys.}}
  \textbf{\bibinfo{volume}{78}}, \bibinfo{pages}{17--85}
  (\bibinfo{year}{2006}).
\newblock \urlprefix\url{http://link.aps.org/doi/10.1103/RevModPhys.78.17}.

\bibitem{Sachdev2007Book}
\bibinfo{author}{Sachdev, S.}
\newblock \emph{\bibinfo{title}{Quantum phase transitions}}
  (\bibinfo{publisher}{Wiley Online Library}, \bibinfo{year}{2007}).

\bibitem{SokoloffWidom1975MagneticOrderingInHe3}
\bibinfo{author}{Sokoloff, J.~B.} \& \bibinfo{author}{Widom, A.}
\newblock \bibinfo{title}{Magnetic ordering in solid helium-3}.
\newblock \emph{\bibinfo{journal}{Phys. Rev. Lett.}}
  \textbf{\bibinfo{volume}{35}}, \bibinfo{pages}{673--676}
  (\bibinfo{year}{1975}).
\newblock \urlprefix\url{http://link.aps.org/doi/10.1103/PhysRevLett.35.673}.

\bibitem{Montambaux1982He3Vacancies}
\bibinfo{author}{Montambaux, G.}, \bibinfo{author}{Heritier, M.} \&
  \bibinfo{author}{Lederer, P.}
\newblock \bibinfo{title}{Vacancies in a quantum crystal of fermions. the spin
  polaron in solid 3 he}.
\newblock \emph{\bibinfo{journal}{Journal of Low Temperature Physics}}
  \textbf{\bibinfo{volume}{47}}, \bibinfo{pages}{39--89}
  (\bibinfo{year}{1982}).

\bibitem{Kumar1985LowTSusceptibilityInSolidHe3}
\bibinfo{author}{Kumar, P.} \& \bibinfo{author}{Sullivan, N.~S.}
\newblock \bibinfo{title}{Anomalous low-temperature susceptibility of solid
  $^{3}\mathrm{He}$ at high molar volumes}.
\newblock \emph{\bibinfo{journal}{Phys. Rev. Lett.}}
  \textbf{\bibinfo{volume}{55}}, \bibinfo{pages}{963--965}
  (\bibinfo{year}{1985}).
\newblock \urlprefix\url{http://link.aps.org/doi/10.1103/PhysRevLett.55.963}.

\bibitem{Kumar1987MagneticPolaronsInHe3}
\bibinfo{author}{Kumar, P.} \& \bibinfo{author}{Sullivan, N.~S.}
\newblock \bibinfo{title}{Magnetic polarons in low-density solid
  $^{3}\mathrm{He}$}.
\newblock \emph{\bibinfo{journal}{Phys. Rev. B}} \textbf{\bibinfo{volume}{35}},
  \bibinfo{pages}{3162--3169} (\bibinfo{year}{1987}).
\newblock \urlprefix\url{http://link.aps.org/doi/10.1103/PhysRevB.35.3162}.

\bibitem{Lebed2008OrganicSuperconductorsBook}
\bibinfo{author}{Lebed, A.~G.}
\newblock \emph{\bibinfo{title}{The physics of organic superconductors and
  conductors}}, vol. \bibinfo{volume}{110} (\bibinfo{publisher}{Springer},
  \bibinfo{year}{2008}).

\bibitem{Tokura1999ColossalMagnetoresistanceManganites}
\bibinfo{author}{Tokura, Y.} \& \bibinfo{author}{Tomioka, Y.}
\newblock \bibinfo{title}{Colossal magnetoresistive manganites}.
\newblock \emph{\bibinfo{journal}{Journal of magnetism and magnetic materials}}
  \textbf{\bibinfo{volume}{200}}, \bibinfo{pages}{1--23}
  (\bibinfo{year}{1999}).

\bibitem{Roder1996ColossalMagnetoresistanceManganites}
\bibinfo{author}{R{\"o}der, H.}, \bibinfo{author}{Zang, J.} \&
  \bibinfo{author}{Bishop, A.~R.}
\newblock \bibinfo{title}{Lattice effects in the colossal-magnetoresistance
  manganites}.
\newblock \emph{\bibinfo{journal}{Physical review letters}}
  \textbf{\bibinfo{volume}{76}}, \bibinfo{pages}{1356} (\bibinfo{year}{1996}).

\bibitem{Greiner2002BosonicMI}
\bibinfo{author}{Greiner, M.}, \bibinfo{author}{Mandel, O.},
  \bibinfo{author}{Esslinger, T.}, \bibinfo{author}{H{\"a}nsch, T.~W.} \&
  \bibinfo{author}{Bloch, I.}
\newblock \bibinfo{title}{Quantum phase transition from a superfluid to a mott
  insulator in a gas of ultracold atoms}.
\newblock \emph{\bibinfo{journal}{nature}} \textbf{\bibinfo{volume}{415}},
  \bibinfo{pages}{39--44} (\bibinfo{year}{2002}).

\bibitem{Greif2016FermionicMI}
\bibinfo{author}{Greif, D.} \emph{et~al.}
\newblock \bibinfo{title}{Site-resolved imaging of a fermionic mott insulator}.
\newblock \emph{\bibinfo{journal}{Science}} \textbf{\bibinfo{volume}{351}},
  \bibinfo{pages}{953--957} (\bibinfo{year}{2016}).

\bibitem{Weitenberg2011SingleSpinAddressing}
\bibinfo{author}{Weitenberg, C.} \emph{et~al.}
\newblock \bibinfo{title}{Single-spin addressing in an atomic mott insulator}.
\newblock \emph{\bibinfo{journal}{Nature}} \textbf{\bibinfo{volume}{471}},
  \bibinfo{pages}{319--324} (\bibinfo{year}{2011}).

\bibitem{Endres2011StringOrder}
\bibinfo{author}{Endres, M.} \emph{et~al.}
\newblock \bibinfo{title}{Observation of correlated particle-hole pairs and
  string order in low-dimensional mott insulators}.
\newblock \emph{\bibinfo{journal}{Science}} \textbf{\bibinfo{volume}{334}},
  \bibinfo{pages}{200--203} (\bibinfo{year}{2011}).

\bibitem{Cheuk2016FermionicMIWithSingleSiteResolution}
\bibinfo{author}{Cheuk, L.~W.} \emph{et~al.}
\newblock \bibinfo{title}{Observation of 2d fermionic mott insulators of k 40
  with single-site resolution}.
\newblock \emph{\bibinfo{journal}{Physical Review Letters}}
  \textbf{\bibinfo{volume}{116}}, \bibinfo{pages}{235301}
  (\bibinfo{year}{2016}).

\bibitem{Hilker2017HiddenAFM}
\bibinfo{author}{Hilker, T.~A.} \emph{et~al.}
\newblock \bibinfo{title}{Revealing hidden antiferromagnetic correlations in
  doped hubbard chains via string correlators}.
\newblock \emph{\bibinfo{journal}{arXiv preprint arXiv:1702.00642}}
  (\bibinfo{year}{2017}).

\bibitem{Feynman1966LecturesOnPhysics}
\bibinfo{author}{Feynman, R.~P.}, \bibinfo{author}{Leighton, R.~B.},
  \bibinfo{author}{Sands, M.} \& \bibinfo{author}{Lindsay, R.~B.}
\newblock \bibinfo{title}{The feynman lectures on physics, vol. 3: Quantum
  mechanics} (\bibinfo{year}{1966}).

\bibitem{Esterling1970DegenerateSpinEnvironment}
\bibinfo{author}{Esterling, D.~M.} \& \bibinfo{author}{Lange, R.~V.}
\newblock \bibinfo{title}{Hubbard model. i. degeneracy in the atomic light}.
\newblock \emph{\bibinfo{journal}{Phys. Rev. B}} \textbf{\bibinfo{volume}{1}},
  \bibinfo{pages}{2231--2237} (\bibinfo{year}{1970}).
\newblock \urlprefix\url{http://link.aps.org/doi/10.1103/PhysRevB.1.2231}.

\bibitem{Handscomb1962MonteCarloBook}
\bibinfo{author}{Handscomb, D.}
\newblock \bibinfo{title}{The monte carlo method in quantum statistical
  mechanics}.
\newblock In \emph{\bibinfo{booktitle}{Mathematical Proceedings of the
  Cambridge Philosophical Society}}, vol.~\bibinfo{volume}{58},
  \bibinfo{pages}{594--598} (\bibinfo{organization}{Cambridge Univ Press},
  \bibinfo{year}{1962}).

\bibitem{BrinkmanRice1970}
\bibinfo{author}{Brinkman, W.~F.} \& \bibinfo{author}{Rice, T.~M.}
\newblock \bibinfo{title}{Single-particle excitations in magnetic insulators}.
\newblock \emph{\bibinfo{journal}{Phys. Rev. B}} \textbf{\bibinfo{volume}{2}},
  \bibinfo{pages}{1324--1338} (\bibinfo{year}{1970}).
\newblock \urlprefix\url{http://link.aps.org/doi/10.1103/PhysRevB.2.1324}.

\bibitem{Trugman1988}
\bibinfo{author}{Trugman, S.~A.}
\newblock \bibinfo{title}{Interaction of holes in a hubbard antiferromagnet and
  high-temperature superconductivity}.
\newblock \emph{\bibinfo{journal}{Phys. Rev. B}} \textbf{\bibinfo{volume}{37}},
  \bibinfo{pages}{1597--1603} (\bibinfo{year}{1988}).
\newblock \urlprefix\url{http://link.aps.org/doi/10.1103/PhysRevB.37.1597}.

\bibitem{Jones1998GroupTheoryBook}
\bibinfo{author}{Jones, H.~F.}
\newblock \emph{\bibinfo{title}{Groups, representations and physics}}
  (\bibinfo{publisher}{CRC Press}, \bibinfo{year}{1998}).

\bibitem{Ogata1990SpinChargeSeparation1DBetheAnsatz}
\bibinfo{author}{Ogata, M.} \& \bibinfo{author}{Shiba, H.}
\newblock \bibinfo{title}{Bethe-ansatz wave function, momentum distribution,
  and spin correlation in the one-dimensional strongly correlated hubbard
  model}.
\newblock \emph{\bibinfo{journal}{Physical Review B}}
  \textbf{\bibinfo{volume}{41}}, \bibinfo{pages}{2326} (\bibinfo{year}{1990}).

\bibitem{Kruis2004SpinChargeSeparationInLuttingerLiquids}
\bibinfo{author}{Kruis, H.}, \bibinfo{author}{McCulloch, I.},
  \bibinfo{author}{Nussinov, Z.} \& \bibinfo{author}{Zaanen, J.}
\newblock \bibinfo{title}{Geometry and the hidden order of luttinger liquids:
  The universality of squeezed space}.
\newblock \emph{\bibinfo{journal}{Physical Review B}}
  \textbf{\bibinfo{volume}{70}}, \bibinfo{pages}{075109}
  (\bibinfo{year}{2004}).

\bibitem{Bulaevski1968}
\bibinfo{author}{Bulaevski, L.}, \bibinfo{author}{Nagaev, E.} \&
  \bibinfo{author}{Khomskii, D.}
\newblock \bibinfo{title}{A new type of auto-localized state of a conduction
  electron in an antiferromagnetic semiconductor}.
\newblock \emph{\bibinfo{journal}{Soviet Journal of Experimental and
  Theoretical Physics}} \textbf{\bibinfo{volume}{27}}, \bibinfo{pages}{836}
  (\bibinfo{year}{1968}).

\bibitem{SchmittRinkVarma1988SpectralFunctionOfTheHoleInAQuantumAFM}
\bibinfo{author}{Schmitt-Rink, S.}, \bibinfo{author}{Varma, C.} \&
  \bibinfo{author}{Ruckenstein, A.}
\newblock \bibinfo{title}{Spectral function of holes in a quantum
  antiferromagnet}.
\newblock \emph{\bibinfo{journal}{Physical review letters}}
  \textbf{\bibinfo{volume}{60}}, \bibinfo{pages}{2793} (\bibinfo{year}{1988}).

\bibitem{Kane1989MotionOfASingleHoleInAQuantumAFM}
\bibinfo{author}{Kane, C.}, \bibinfo{author}{Lee, P.} \& \bibinfo{author}{Read,
  N.}
\newblock \bibinfo{title}{Motion of a single hole in a quantum
  antiferromagnet}.
\newblock \emph{\bibinfo{journal}{Physical Review B}}
  \textbf{\bibinfo{volume}{39}}, \bibinfo{pages}{6880} (\bibinfo{year}{1989}).

\bibitem{Mierzejewski2011NonEqQDynamics}
\bibinfo{author}{Mierzejewski, M.}, \bibinfo{author}{Vidmar, L.},
  \bibinfo{author}{Bon\ifmmode~\check{c}\else \v{c}\fi{}a, J.} \&
  \bibinfo{author}{Prelov\ifmmode~\check{s}\else \v{s}\fi{}ek, P.}
\newblock \bibinfo{title}{Nonequilibrium quantum dynamics of a charge carrier
  doped into a mott insulator}.
\newblock \emph{\bibinfo{journal}{Phys. Rev. Lett.}}
  \textbf{\bibinfo{volume}{106}}, \bibinfo{pages}{196401}
  (\bibinfo{year}{2011}).
\newblock
  \urlprefix\url{http://link.aps.org/doi/10.1103/PhysRevLett.106.196401}.

\bibitem{Bonca2012HolePairDynamics}
\bibinfo{author}{Bon\ifmmode~\check{c}\else \v{c}\fi{}a, J.},
  \bibinfo{author}{Mierzejewski, M.} \& \bibinfo{author}{Vidmar, L.}
\newblock \bibinfo{title}{Nonequilibrium propagation and decay of a bound pair
  in driven $t\mathrm{\text{\ensuremath{-}}}j$ models}.
\newblock \emph{\bibinfo{journal}{Phys. Rev. Lett.}}
  \textbf{\bibinfo{volume}{109}}, \bibinfo{pages}{156404}
  (\bibinfo{year}{2012}).
\newblock
  \urlprefix\url{http://link.aps.org/doi/10.1103/PhysRevLett.109.156404}.

\bibitem{DalConte2015RetardedInteractionInCuprates}
\bibinfo{author}{Dal~Conte, S.} \emph{et~al.}
\newblock \bibinfo{title}{Snapshots of the retarded interaction of charge
  carriers with ultrafast fluctuations in cuprates}.
\newblock \emph{\bibinfo{journal}{Nature Physics}}
  \textbf{\bibinfo{volume}{11}}, \bibinfo{pages}{421--426}
  (\bibinfo{year}{2015}).

\bibitem{Shastry1990NagaokaInsability}
\bibinfo{author}{Shastry, B.~S.}, \bibinfo{author}{Krishnamurthy, H.~R.} \&
  \bibinfo{author}{Anderson, P.~W.}
\newblock \bibinfo{title}{Instability of the nagaoka ferromagnetic state of the
  u=\ensuremath{\infty} hubbard model}.
\newblock \emph{\bibinfo{journal}{Phys. Rev. B}} \textbf{\bibinfo{volume}{41}},
  \bibinfo{pages}{2375--2379} (\bibinfo{year}{1990}).
\newblock \urlprefix\url{http://link.aps.org/doi/10.1103/PhysRevB.41.2375}.

\bibitem{Sandvik1991StochasticSeriesExpansion}
\bibinfo{author}{Sandvik, A.~W.} \& \bibinfo{author}{Kurkij\"arvi, J.}
\newblock \bibinfo{title}{Quantum monte carlo simulation method for spin
  systems}.
\newblock \emph{\bibinfo{journal}{Phys. Rev. B}} \textbf{\bibinfo{volume}{43}},
  \bibinfo{pages}{5950--5961} (\bibinfo{year}{1991}).
\newblock \urlprefix\url{http://link.aps.org/doi/10.1103/PhysRevB.43.5950}.

\end{thebibliography}
\bibliographystyle{naturemag}

\end{document}